\documentclass[12pt,preprint]{aastex}

\begin{document}

\title{Orientation Angles of a Pulsar's Polarization Vector}
\author{Mark M. McKinnon}
\affil{National Radio Astronomy Observatory,\altaffilmark{1}\altaffiltext{1}
{The National Radio Astronomy Observatory is a facility of the National 
Science Foundation operated under cooperative agreement by Associated 
Universities, Inc.} Socorro, NM \ \ 87801\ \ USA}

\begin{abstract}

  A statistical model of the polarization of pulsar radio emission is used
to derive the general statistics of a polarization vector's orientation 
angles. The theoretical distributions are compared with orientation angle 
histograms computed from single-pulse, polarization observations of PSR 
B2020+28. The favorable agreement between the theoretical and measured 
distributions lends support to the underlying assumptions of the statistical 
model, and demonstrates, like recent work on other pulsars, that the 
handedness of circular polarization is associated with the radiation's 
orthogonally polarized modes. Comprehensive directional statistics of the 
vector's orientation angles are also derived, and are shown to follow the 
Watson bipolar and Fisher distributions in its limiting forms. 

\end{abstract}


\section{INTRODUCTION}

  Apart from the classic sweep of the polarization position angle across 
a pulsar's average profile, the orthogonal polarization modes (OPM) are 
perhaps the most striking feature in the polarization of pulsar radio emission.
The modes appear in bimodal histograms of polarization position angle as two 
concentrations of data points separated by about $90^\circ$ (e.g. Stinebring 
et al. 1984). The modes are elliptically as well as orthogonally polarized, so 
that the polarization states of the modes reside at antipodal points on the 
Poincar\'e sphere (Cordes et al. 1978). OPM are thought to arise from wave 
propagation effects in the magnetized plasma above the pulsar's polar cap 
(e.g. Allen \& Melrose 1982; Barnard \& Arons 1986). The simultaneous 
interaction of the modes may be responsible for the depolarization of the 
emission at high radio frequency (Manchester et al. 1975; Morris et al. 1981).

  The notion that OPM are elliptically polarized suggests that the emission's 
linear and circular polarization should be analyzed together, instead of 
individually, for a proper interpretation of pulsar polarization measurements 
(McKinnon 2003; hereafter M03). Comprehensive methods have been developed 
recently for displaying and analysing the polarization data in this way. 
For example, Karastergiou et al. (2003) plotted the measured orientation 
angles of a pulsar's polarization vector in a Hammer-Aitoff projection, 
confirming the association of circular polarization handedness with OPM in 
PSR B1133+16. Similarly, Edwards \& Stappers (2004; hereafter ES04) displayed 
the vector's orientation angles in a Lambert equal-area projection to document 
and explore the intriguing deviations from mode orthogonality in PSR B0329+54. 
They also showed that the circular polarization in the pulsar is generally 
consistent with an origin in elliptically polarized OPM. McKinnon \& Stinebring 
(1998, 2000) developed a statistical model of superposed OPM that predicts that 
polarization measurements at a given pulse longitude should form a single, 
prolate ellipsoid in the three-dimensional space defined by the Stokes parameters 
Q, U, and V. The prediction is generally supported by the analyses conducted by 
McKinnon (2004; hereafter M04) and ES04, who independently developed eigenvalue 
techniques to quantify the dimensions and orientations of the ellipsoids. While 
the eigenvalue analyses and the projections of polarization vector angles are 
generally supportive of the superposed OPM hypothesis, it has yet to be 
demonstrated that the observed orientation angles are consistent with those 
predicted by the OPM statistical model. The statistics of the orientation angles 
derived in M03 are limited in their applicability, primarily because they are 
restricted to the specific case of OPM with equal polarization amplitudes.

  The objectives of this paper are to derive a general distribution for a 
polarization vector's orientation angles by expanding the work of M03 and to 
compare the theoretical distribution with the distributions of observed 
orientation angles. The derivation is presented in \S\ref{sec:angles}, and 
the comparison is made in \S\ref{sec:compare}. Conclusions are discussed in 
\S\ref{sec:conclude}. 

\section{THEORETICAL DISTRIBUTIONS OF ORIENTATION ANGLES}
\label{sec:angles}

  The joint probability density of a polarization vector's amplitude,
colatitude, and longitude must be determined from the Stokes parameters to 
calculate the individual distributions of the orientation angles. From the 
statistical model of pulsar polarization (M03), the equations for the measured 
Stokes parameters are 
\begin{equation}
{\rm Q} = \sin\theta_o\cos\phi_o (X_1 - X_2) + X_{\rm N,Q},
\end{equation}
\begin{equation}
{\rm U} = \sin\theta_o\sin\phi_o (X_1 - X_2) + X_{\rm N,U},
\end{equation}
\begin{equation}
{\rm V} = \cos\theta_o (X_1 - X_2) + X_{\rm N,V}.
\end{equation}

\noindent The random variables $X_1$ and $X_2$ represent the amplitudes of the 
mode polarization vectors. They have means $\mu_1$ and $\mu_2$ and standard 
deviations $\sigma_1$ and $\sigma_2$. In general, $X_1$ and $X_2$ can be 
correlated random variables, but they are assumed to be independent in what
follows. The Gaussian random variable $X_{\rm N}$ accounts for the instrumental 
noise. It has a zero mean and a standard deviation $\sigma_{\rm N}$. The fixed 
angles $\theta_o$ and $\phi_o$ are the colatitude and longitude, respectively, 
of the diagonal in the Poincar\'e sphere along which the fluctuations in 
pulsar-intrinsic polarization occur. The angles correspond to twice the 
ellipticity and twice the position angle, respectively, of the polarization 
vector (i.e.  $\theta_o = 2\chi_o$ and $\phi_o = 2\psi_o$ in the nomenclature 
of ES04). As discussed in M04, the emission may contain additional components, 
such as randomly polarized radiation (RPR). For the purposes of this paper, 
these components are neglected, or alternatively, assumed to be absorbed 
in the $X_{\rm N}$ terms provided that their fluctuations are Gaussian with a 
zero mean and that their Q-U-V covariance matrix is a scaled identity matrix.

  The basic shape of a data point cluster formed by the equations for the 
Stokes parameters is a prolate ellipsoid. The orientation of the ellipsoid's 
major axis is determined by $\theta_o$ and $\phi_o$. To simplify the derivation 
of the joint probability density of a polarization vector's amplitude and 
orientation, it is convenient to specify $\theta_o=0$ so that the ellipsoid 
is symmetric about the z-axis in a conventional Cartesian coordinate system. 
By virtue of this symmetry, the longitude of the polarization vector is 
uniformly distributed over the interval $0\le\phi < 2\pi$, and the longitude 
and colatitude distributions are statistically independent. Since the ellipsoid 
retains its shape on rotation, measured distributions of colatitude and longitude 
can be compared with the theoretical result by determining the orientation 
of the measured ellipsoid's major axis, rotating the Q-U-V data points that
form the ellipsoid via matrix multiplication so that the major axis of the 
rotated ellipsoid is aligned with the z-axis, and computing histograms of the 
rotated values of colatitude and longitude. 

  When $\theta_o=0$ and when the mode polarization amplitudes, $X_1$ and $X_2$, 
are independent, Gaussian random variables, the joint probability density of 
the measured polarization vector's amplitude and orientation is

\begin{equation}
f(r,\theta,\phi) = {r^2\over{\sigma_N^3}}{\sin\theta\over{(2\pi)^{3/2}}}
                   {1\over{(1+\rho^2)^{1/2}}}
             \exp\Biggl[{-r^2(1+\rho^2\sin^2\theta)+ 2r\mu\cos\theta -\mu^2\over
                    {2\sigma_N^2(1+\rho^2)}}\Biggr].
\label{eqn:jdc}
\end{equation}

\noindent The parameter $\mu = \mu_1-\mu_2$ is the mean value of the pulsar
intrinsic polarization, and the parameter 
$\rho = (\sigma_1^2+\sigma_2^2)^{1/2}/\sigma_{\rm N}$ is a measure of the 
intrinsic polarization fluctuations relative to the instrumental noise. 
Following the procedure outlined in M03, the joint probability density can 
be used to find the distribution of the polarization vector's amplitude.

\begin{equation}
f(r) = {r^2\over{\sigma_N^3}} {1\over{[2\pi(1+\rho^2)]^{1/2}}}
       \exp\Biggl[-{(\rho^2r^2+\mu^2)\over {2\rho^2\sigma_N^2}}\Biggr]
       \int_{\mu/r\rho^2-1}^{\mu/r\rho^2+1}
       \exp\Biggl[{\rho^2r^2x^2\over{2(1+\rho^2)\sigma_N^2}}\Biggr]dx
\label{eqn:ampcm}
\end{equation}

\noindent Similarly, the vector's colatitude distribution is
 
\begin{eqnarray}
f(\theta) & = & {\sin\theta\over{2}} {(1+\rho^2)\over{(1+\rho^2\sin^2\theta)^{3/2}}}
 \Biggl\{\exp{\Biggl[-{s^2\sin^2\theta\over {2(1+\rho^2\sin^2\theta)}} \Biggr]}
 \Biggl[1+{\rm erf}\Biggl({y\over{\sqrt{2}}}\Biggr)\Biggr](1 + y^2) \nonumber \\
 & + & y\sqrt{{2\over{\pi}}}\exp{\Biggl[-{s^2\over{2(1+\rho^2)}}\Biggr]}\Biggr\},
\label{eqn:gencolat}
\end{eqnarray}

\noindent where $y$ is a function of $\rho$, $s$, and $\theta$ given by

\begin{equation}
y={s\cos\theta\over{[(1 + \rho^2\sin^2\theta)(1 + \rho^2)]^{1/2}}}.
\end{equation}

\begin{figure}
\plotone{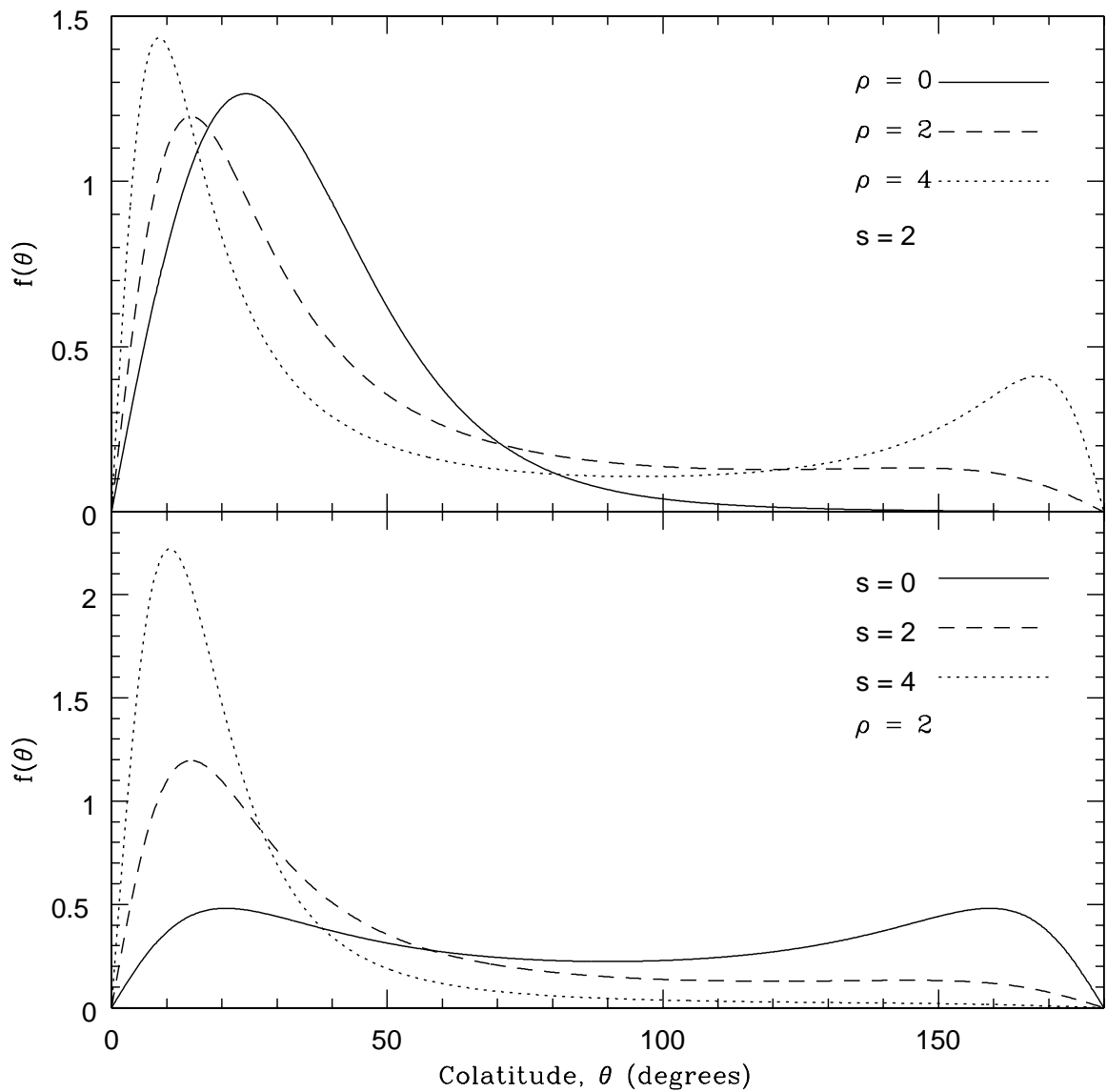}
\caption{Distributions of a polarization vector's colatitude, $\theta$ 
(eqn.~\ref{eqn:gencolat}). The top panel shows colatitude distributions for a 
fixed signal-to-noise ratio in polarization, $s$, with varying mode fluctuation 
ratio, $\rho$. The bottom panel shows colatitude distributions for fixed $\rho$ 
and varying $s$.}
\label{fig:colat}
\end{figure}

\noindent The colatitude distribution is characterized by two free parameters; 
the signal-to-noise ratio in polarization, $s=\mu/\sigma_{\rm N}$, and the mode 
fluctuation ratio, $\rho$. The distribution is shown in Figure~\ref{fig:colat} 
for different values of $s$ and $\rho$. The distribution is distinctly bimodal 
for large values of $\rho$, and becomes more sharply peaked at large values of 
$s$. Equation~\ref{eqn:gencolat} generalizes the case-specific colatitude 
distributions for $s=0$ and $\rho=0$ derived in M03. 

  The conditional density of the polarization vector's colatitude, or the 
colatitude density at a fixed value of the polarization amplitude, $r_o$, is 

\begin{equation}
f(\theta |r_o) = 2\pi{f(r_o,\theta,\phi)\over{f(r_o)}}
                    = {\sin\theta\over{w(\kappa,\gamma)}}
                       \exp[\kappa(\gamma + \cos\theta)^2],
\label{eqn:comp}
\end{equation}

\noindent where $w(\kappa,\gamma)$ is a normalization factor given by 

\begin{equation}
w(\kappa,\gamma) = \int_{\gamma - 1}^{\gamma + 1}\exp(\kappa x^2)dx.
\end{equation}

\begin{figure}
\plotone{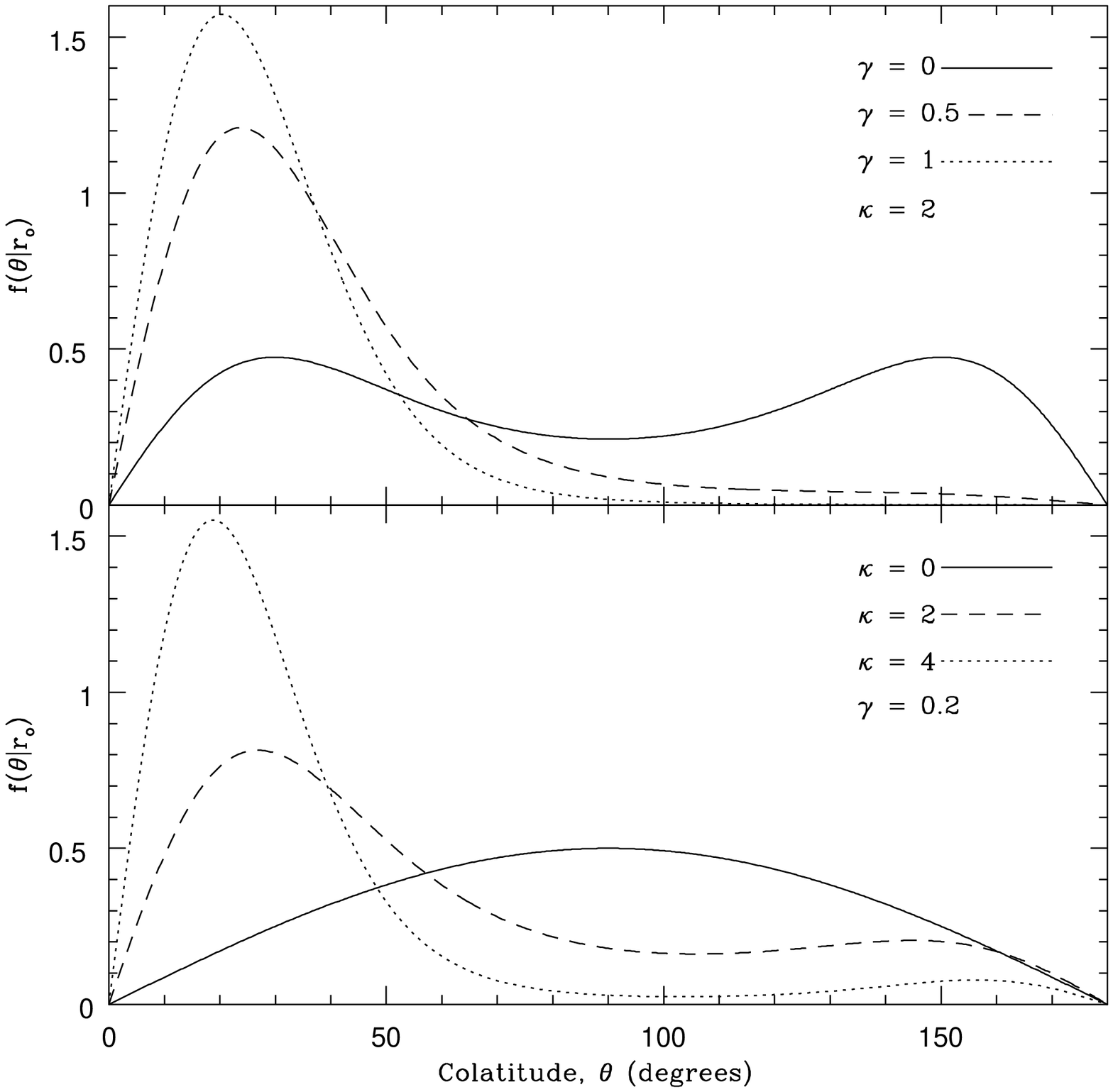}
\caption{Conditional density of a polarization vector's colatitude
(eqn.~\ref{eqn:comp}). The distributions are shown for different 
concentration parameters, $\kappa$, and symmetry parameters, $\gamma$. The 
top panel shows distributions for varying $\gamma$ with $\kappa$ fixed. The 
bottom panel shows distributions for varying $\kappa$ with $\gamma$ fixed.}
\label{fig:comp}
\end{figure}

\noindent The colatitude conditional density is characterized by two 
dimensionless quantities; a concentration parameter given by 
$\kappa=r_o^2\rho^2/2(1+\rho^2)\sigma_N^2$ and a symmetry parameter given by 
$\gamma = \mu/r_o\rho^2$. Examples of the conditional density are shown in 
Figure~\ref{fig:comp}. The peaks in the distribution narrow as $\kappa$ 
increases. When $\gamma=0$ (e.g. when $\mu = 0$), the conditional density 
becomes the Watson bipolar distribution (eqn. 38 of M03), which is always 
symmetric about $\theta=\pi/2$. As $\gamma$ increases, the conditional 
density is dominated by the strong polarization mode and peaks at small values 
of $\theta$. The parameters $\kappa$ and $\gamma$ are not entirely independent. 
For example, when $\rho\ll 1$, $\kappa$ becomes small and $\gamma$ becomes 
very large, and the conditional density becomes the Fisher distribution with a 
concentration parameter of $\kappa_f=2\gamma\kappa\simeq \mu r_o/\sigma_{\rm N}^2$ 
(see eqn. 37 of M03). Statistical treatments of directional data (e.g.  Fisher 
et al. 1987) appear to consider the Watson and Fisher distributions as 
completely unrelated. However, the analysis presented here clearly shows that 
the two distributions are special cases of the more comprehensive conditional 
density given by equation~\ref{eqn:comp}.

  Figure~\ref{fig:colat} shows that the peaks in the colatitude distribution 
do not occur at the poles of the Poincar\'e sphere. In fact, both the locations 
and the widths of the peaks depend upon the values of $s$ and $\rho$. The 
distribution gives the illusion that the data points in the Q-U-V cluster form 
annuli around the poles. The illusion arises from the computation of the 
colatitude distribution, which does not preserve the density of data points in 
the cluster (see also the discussion of equal-area projections in ES04). The 
polar concentration of data points is retained in the distribution of 
$u = \cos\theta$, where the locations of the OPM peaks occur at $\theta=0,\pi$ 
and only the widths of the peaks are affected by $s$ and $\rho$. The 
distribution of $u$ can be computed directly from the distribution of $\theta$ 
given by equation~\ref{eqn:gencolat}. The distribution is 

\begin{eqnarray}
f(u) & = & {1\over{2}} {(1+\rho^2)\over{[1+\rho^2(1-u^2)]^{3/2}}}
 \Biggl\{\exp{\Biggl[-{s^2(1-u^2)\over {2[1+\rho^2(1-u^2)]}} \Biggr]}
 \Biggl[1+{\rm erf}\Biggl({\bar y\over{\sqrt{2}}}\Biggr)\Biggr]
 (1+\bar y^2) \nonumber \\
 & + & \bar y\sqrt{{2\over{\pi}}}\exp{\Biggl[-{s^2\over{2(1+\rho^2)}}\Biggr]}\Biggr\},
\label{eqn:coscolat}
\end{eqnarray}

\noindent where $\bar y$ is given by

\begin{equation}
\bar y = {su\over{\{[1 + \rho^2(1-u^2)](1 + \rho^2)\}^{1/2}}}.
\end{equation}

\noindent The conditional density of $u$ can be derived from
equation~\ref{eqn:comp}.

\begin{equation}
f(u |r_o) = {\exp[\kappa(\gamma + u)^2]\over{w(\kappa,\gamma)}}
\label{eqn:ugen}
\end{equation}

\section{DISTRIBUTION COMPARISON}
\label{sec:compare}

  To compare the measured distributions of a polarization vector's 
colatitude and longitude with the analytical results, the Q-U-V data 
points comprising the polarization ellipsoid must be rotated so that 
the ellipsoid's major axis is aligned with the z-axis. From Fisher
et al. (1987), the matrix multiplication operation that performs the 
rotation is

\begin{equation}
\left[\matrix{\cos\theta_o\cos\phi_o& \cos\theta_o\sin\phi_o& -\sin\theta_o\cr
        -\sin\phi_o& \cos\phi_o& 0\cr
         \sin\theta_o\cos\phi_o& \sin\theta_o\sin\phi_o& \cos\theta_o\cr}\right]
\left[\matrix{Q_o\cr U_o\cr V_o\cr}\right] = \left[\matrix{0\cr 0\cr \mu\cr}\right].
\label{eqn:rotate}
\end{equation}

  Care must be exercised in choosing appropriate estimates of the ellipsoid's
orientation angles, $\theta_o$ and $\phi_o$, before the rotation is performed. 
The angles could be estimated from the mean values of the Stokes parameters at 
the pulse longitude of interest from

\begin{equation}
\phi_o = \arctan\Biggl({\langle U\rangle\over{\langle Q\rangle}}\Biggr),
\label{eqn:phi}
\end{equation}

\begin{equation}
\theta_o = \arccos\Biggl[{\langle V\rangle\over{
(\langle Q\rangle^2 + \langle U\rangle^2 + \langle V\rangle^2)^{1/2}}}\Biggr],
\label{eqn:theta}
\end{equation}

\noindent which would suffice for locations within the pulse where the 
polarization signal-to-noise ratio is high, $s\gg 0$. But at locations where 
$s\simeq 0$, the mean Stokes parameters will be small, and consequently 
equations~\ref{eqn:phi} and~\ref{eqn:theta} will produce poor estimates 
of $\phi_o$ and $\theta_o$. Alternatively, one could compute the covariances
of the Stokes parameters, and the directional cosines of the principal 
eigenvector of the 3x3 covariance matrix could be used to estimate $\phi_o$ 
and $\theta_o$. The polarization ellipsoid can be highly elongated where OPM 
occurs, and the directional cosines of the principal eigenvector can be well 
determined, even where $s=0$, because its eigenvalue, 
$\tau_{11}=\sigma_N^2(1+\rho^2)$ (M04), is a strong function of the mode 
fluctuation ratio, $\rho$. However, this method produces poor estimates of 
the orientation angles at locations where $\rho\simeq 0$, even if $s\gg 0$, 
because the eigenvalues in this case are equal 
($\tau_{11}=\tau_{22}=\tau_{33}= \sigma_N^2$), and thus the eigenvectors are 
not unique. The best estimate of the ellipsoid's orientation angles, and the 
one used in the analysis presented below, comes from the directional cosines 
of the principal eigenvector determined from the second moments of the Stokes 
parameters. The eigenvalue of the second moment principal eigenvector is 
$\tau_{11}=\sigma_N^2(1+s^2+\rho^2)$, and the directional cosines of the 
eigenvector will be well determined when either $s\simeq 0$ or $\rho\simeq 0$. 
When both $s$ and $\rho$ are small, no polarized signal is present, the 
polarization ellipsoid is a spheroid centered on the origin of Q-U-V space, 
and there is no need to rotate the spheroid (i.e. the distribution of data 
points is isotropic).

  Histograms of the measured values of colatitude and longitude were 
computed at different locations within the pulse of PSR B2020+28 using 
the 1404 MHz single pulse, polarization observations of Stinebring 
et al. (1984) as follows. From the multiple measurements of the Stokes 
parameters taken at each location, the second moments of the Stokes 
parameters were computed and used to find the directional cosines of 
the polarization ellipsoid's principal eigenvector (major axis). The 
directional cosines were then used to compute the orientation angles 
of the major axis. The resulting values of $\theta_o$ and $\phi_o$ were 
used in equation~\ref{eqn:rotate} to rotate the Q-U-V data points that 
formed the polarization ellipsoid. Finally, the rotated Stokes parameters 
were used to calculate corresponding values of colatitude and longitude, 
which were binned into 50 intervals spread over the appropriate range of 
angles to form the histograms.

\begin{figure}
\plotone{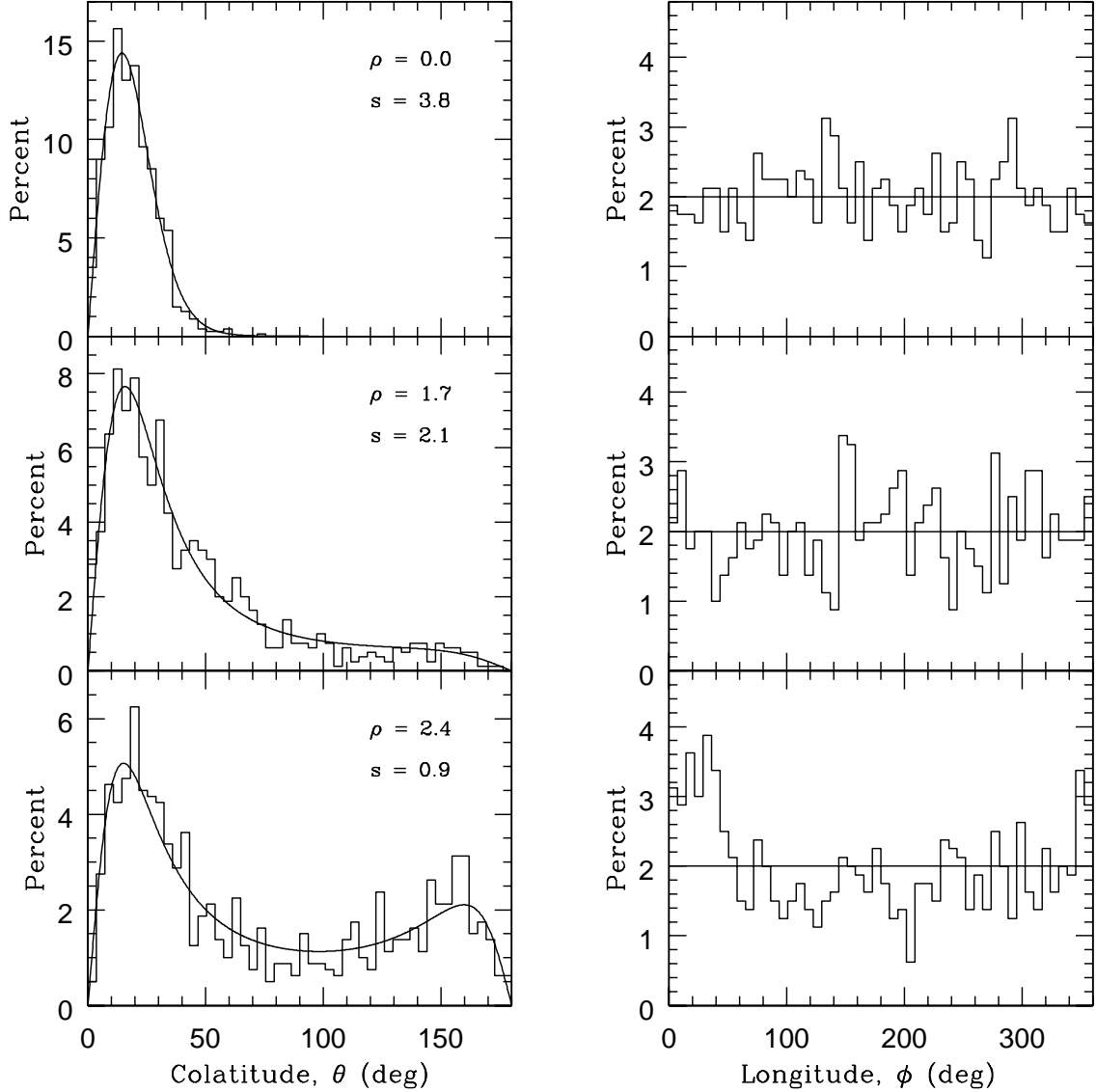}
\caption{Comparison between the measured and theoretical distributions 
of a polarization vector's colatitude and longitude at three different 
locations in the pulse of PSR B2020+28. The longitude histograms are 
generally consisent with the uniform distributions predicted by the 
statistical OPM model, as indicated by the solid horizontal line. The 
smooth curves through the colatitude histograms are best fits to 
equation~\ref{eqn:gencolat}. The fitted values of signal-to-noise ratio, $s$, 
and mode fluctuation ratio, $\rho$, are annotated in each colatitude 
histogram panel.}
\label{fig:6hist}
\end{figure}

  The histograms at three different locations in the pulse are compared 
with the theoretical distributions of the polarization vector's colatitude 
and longitude in Figure~\ref{fig:6hist}. The colatitude distributions are 
shown in the left column of panels in the figure, and their corresponding 
longitude distributions are shown in the right column of panels. The solid 
horizontal line in each of the longitude histogram panels represents the 
uniform distribution expected from the OPM statistical model. The generally
good agreement between the theoretical and measured longitude distributions, 
combined with the observation that the ratios of the minor axes of the Q-U-V 
data point clusters are no larger than 1.09 at these locations within the 
pulse (M04), suggests that the polarization ellipsoids are rotationally 
symmetric about their major axes. The observed symmetry of the ellipsoids 
also suggests that the longitude and colatitude of the polarization vector 
are statistically independent, as predicted by the OPM statistical model. 
However, the symmetry does not rule out the possibility that $\phi$ and 
$\theta$ can be correlated or dependent upon one another. Indeed, the 
deviations from orthogonality documented in ES04 and M04 are evidence that 
the two angles can be correlated. The smooth, solid curves shown in the 
colatitude histogram panels represent the best fits of the data to 
equation~\ref{eqn:gencolat}. The best fit values of $\rho$ and $s$ are 
annotated in each panel. The excellent fits show that 
equation~\ref{eqn:gencolat} is capable of representing a broad range of 
polarization behavior. Furthermore, these fits and the measurements of 
cluster dimensions by M04 together imply that the clusters aren't just 
any ellipsoids, but the ellipsoids expected from the OPM statistical model, 
suggesting quantitative, in addition to qualitative, consistency between 
the model and the data. The data used to construct the histograms had 
significant circular polarization, and the ellipsoidal shape of the Q-U-V 
data point clusters illustrates the association of circular polarization 
handedness with OPM in PSR 2020+28, similar to what was found in PSR 
1133+16 by Karastergiou et al. (2003) and in PSR B0329+54 by ES04. Only 
one polarization mode is apparent ($\rho = 0$) in the colatitude histogram 
shown in the top left panel of the figure. In the lower panels, the other 
polarization mode becomes more obvious as indicated by the bimodal 
histograms and the increasing value of $\rho$. Additionally, the values of 
$s$ in the panels decrease as $\rho$ increases, which is to be expected 
since the simultaneous interaction of the orthogonal modes depolarizes the 
emission. The fitted values of $s$ are low when compared to what is expected 
from the mean polarization and the off-pulse instrumental noise. However, 
the sizes of the ellipsoids' minor axes, and thus the effective noise of 
the observations, are much larger than the off-pulse noise, causing the 
lower-than-expected values of $s$. An emission component such as RPR could 
contribute to the large effective noise (M04, ES04). 

\begin{figure}
\plotone{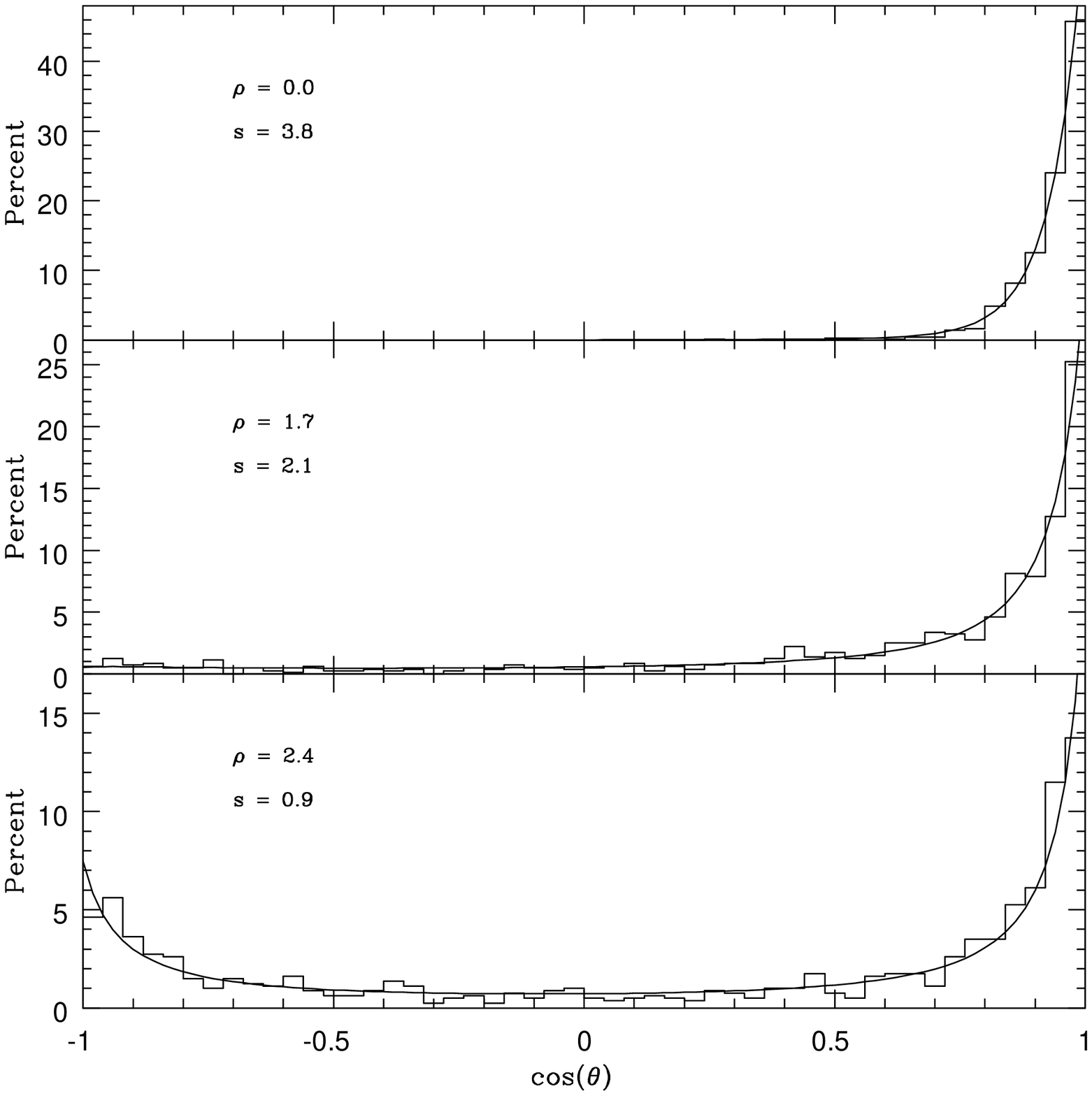}
\caption{Comparison between the measured and theoretical distributions 
of cosine colatitude, $\cos\theta$, at the three locations in PSR B2020+28.}

\label{fig:coscolat}
\end{figure}

  The measured and theoretical distributions of $\cos\theta$ are compared in 
Figure~\ref{fig:coscolat}. The data shown in each panel of the figure are the 
same data used in the corresponding panel of the left column of 
Figure~\ref{fig:6hist}. The histogram in each panel is the measured distribution 
of $\cos\theta$. The smooth curve is the theoretical distribution computed from 
equation~\ref{eqn:coscolat} using the same values of $\rho$ and $s$ as in
Figure~\ref{fig:6hist}. In all panels, the data are concentrated at one or 
both of the poles. The agreement between the theoretical and measured 
distributions is excellent.

\section{DISCUSSION AND CONCLUSIONS}
\label{sec:conclude}

  The statistical model of pulsar polarization developed by McKinnon \&
Stinebring (1998) predicts that multiple measurements of the variable Stokes 
parameters at a fixed pulse longitude will form a single, prolate ellipsoid in 
Poincar\'e space.  The analyses presented in ES04 and M04 show that Q-U-V data 
point clusters in PSR B0329+54, PSR B1929+10, and PSR B2020+28 are indeed 
prolate ellipsoids. In this paper, the favorable comparison between the 
theoretical and measured distributions of polarization vector colatitude and 
longitude in PSR B2020+28 suggest that the data point clusters are consistent 
with the ellipsoid expected from the model. This implies quantitative 
consistency between the model and the data, and supports the fundamental, 
underlying assumption of the statistical model which is the observed 
polarization is determined by the simultaneous interaction of two, 
orthogonally polarized modes.

  The analysis of orientation angles presented here and the observations and
analyses of Cordes et al. (1978), Karastergiou et al. (2003), and ES04 clearly 
demonstrate the association of circular polarization handedness with OPM, 
confirming that the modes are elliptically polarized and occupy antipodal 
points on the Poincar\'e sphere. This result could be interpreted to conclude 
that the circular polarization is formed by the appropriate phasing of the 
naturally occurring, linearly polarized modes through mode coupling (e.g. 
Lyubarskii \& Petrova 1998; Petrova 2001) as the radiation propagates through 
the pulsar's magnetospheric plasma. If this were the case, we would only 
observe circular polarization in parts of the pulse where both modes were 
found. However, there are clear examples of circular polarization where only 
one mode occurs (e.g. top left panel of Fig.~\ref{fig:6hist} and Fig. 3 of 
M04). So, unless one mode just so happens to be completely \lq\lq coupled" 
in the process of creating circular polarization, other propagation effects, 
such as the gyrotropy of the magnetospheric plasma, or cyclotron absorption 
(Melrose 2003) must be responsible for the production of circular polarization 
in the radio emission.

  Karastergiou et al. (2003) and ES04 displayed the orientation angles of 
the polarization vector at a fixed pulse longitude using Hammer-Aitoff and 
Lambert equal-area projections. The colatitude and longitude histograms 
computed in \S\ref{sec:compare} are simply another method for conveying 
the same information. Both the projections and the histograms effectively 
illustrate the fluctuations in the colatitude and longitude of the emission's 
polarization vector. 

  A comprehensive version of the conditional density for the polarization 
vector's orientation angles was derived. In its limiting forms, the 
conditional density follows the Watson bipolar distribution when the
polarization amplitude is small ($\mu=\gamma=0$) and follows the Fisher 
distribution when the polarization fluctuations are small compared to the 
effective noise ($\rho\ll 1$). Since both the Watson and Fisher distributions 
have many applications in the statistical analysis of scientific data (M03), 
the comprehensive version of the conditional density given by
equation~\ref{eqn:comp} may have more widespread applications.

  The small values of polarization signal-to-noise ratio found in the 
orientation angle analysis and the inflated Q-U-V data point clusters found 
by M04 and ES04 are reminders that the effective noise of single pulse, 
polarization observations is much larger than the noise measured off the 
pulse. The additional noise may be due to RPR. The colatitude distribution
(eqn.~\ref{eqn:gencolat}) was derived assuming Gaussian fluctuations in mode 
polarization amplitudes, and it is remarkable that the equation fits the data 
so well, regardless of the actual type of pulsar-intrinsic polarization 
fluctuations. Perhaps the good match between equation and data can be attributed 
to the idea that the observed fluctuations are heavily influenced by the large, 
effective Gaussian noise. The caution to interject here is that we are sifting 
through a large effective noise component in analyzing these data, irrespective 
of how low the off-pulse noise might be. Consequently, it is important to 
account for the effective noise in statistical models so that the data can be 
properly interpreted.

\acknowledgements
Again, I am indebted to Dan Stinebring for his continued generosity in 
providing the data used in the analysis. I thank an anonymous referee
for comments that improved the manuscript.

\vfill\eject

\end{document}